\documentclass[conference]{IEEEtran}
\IEEEoverridecommandlockouts
\usepackage{cite, verbatim}
\usepackage{amsmath,amssymb,amsfonts}
\usepackage{algorithmic}
\usepackage{graphicx}
\usepackage{textcomp}
\usepackage{xcolor}
\usepackage{tikz}
\usepackage{amsmath}
\usepackage{hyperref}
\usepackage{balance}
\newenvironment{formula}{\begin{equation}\begin{aligned}}{\end{aligned}\end{equation}}
\def\BibTeX{{\rm B\kern-.05em{\sc i\kern-.025em b}\kern-.08em
    T\kern-.1667em\lower.7ex\hbox{E}\kern-.125emX}}
\begin{document}

\title{End-to-end Audio Deepfake Detection from RAW Waveforms: a RawNet-Based Approach with Cross-Dataset Evaluation}

\author{
\IEEEauthorblockN{Andrea Di Pierno\IEEEauthorrefmark{1}\IEEEauthorrefmark{2}, Luca Guarnera\IEEEauthorrefmark{2}, Dario Allegra\IEEEauthorrefmark{2}, Sebastiano Battiato\IEEEauthorrefmark{2}}
\IEEEauthorblockA{\IEEEauthorrefmark{1}IMT School of Advanced Studies Lucca, Lucca, Italy}
\IEEEauthorblockA{\IEEEauthorrefmark{2}Department of Mathematics and Computer Science,
University of Catania, Catania, Italy \\
andrea.dipierno@phd.unict.it,
\{luca.guarnera, dario.allegra, sebastiano.battiato\}@unict.it}
}

\maketitle

\begin{abstract}
Audio deepfakes represent a growing threat to digital security and trust, leveraging advanced generative models to produce synthetic speech that closely mimics real human voices. Detecting such manipulations is especially challenging under open-world conditions, where spoofing methods encountered during testing may differ from those seen during training.
In this work, we propose an end-to-end deep learning framework for audio deepfake detection that operates directly on raw waveforms. Our model, \textit{RawNetLite}, is a lightweight convolutional-recurrent architecture designed to capture both spectral and temporal features without handcrafted preprocessing. To enhance robustness, we introduce a training strategy that combines data from multiple domains and adopts Focal Loss to emphasize difficult or ambiguous samples. We further demonstrate that incorporating codec-based manipulations and applying waveform-level audio augmentations (e.g., pitch shifting, noise, and time stretching) leads to significant generalization improvements under realistic acoustic conditions.
The proposed model achieves over 99.7\% F1 and 0.25\% EER on in-domain data (FakeOrReal), and up to 83.4\% F1 with 16.4\% EER on a challenging out-of-distribution test set (AVSpoof2021 + CodecFake). These findings highlight the importance of diverse training data, tailored objective functions and audio augmentations in building resilient and generalizable audio forgery detectors.
Code and pretrained models are available at \href{https://iplab.dmi.unict.it/mfs/Deepfakes/PaperRawNet2025/}{https://iplab.dmi.unict.it/mfs/Deepfakes/PaperRawNet2025/}.
\end{abstract}


\thanks{This study has been partially supported by SERICS (PE00000014) under the MUR National Recovery and Resilience Plan funded by the European Union - NextGenerationEU.}

\section{Introduction}
\label{sec:intro}

Deepfake technologies have rapidly evolved in recent years, enabling the generation of synthetic content that is visually or audibly indistinguishable from real data. While most attention has focused on visual deepfakes—such as face-swapped videos or manipulated facial expressions~\cite{amerini2025deepfake, tolosana2020deepfakes, mirsky2021creation}—advances in generative modeling have also enabled highly convincing \textit{audio} deepfakes~\cite{yi2023audio}. These include text-to-speech (TTS) and voice conversion (VC) systems capable of replicating a speaker’s voice with remarkable fidelity, raising growing concerns over impersonation, fraud, and misinformation in domains such as finance, politics, and social media.

While audio synthesis brings clear benefits for accessibility, entertainment, and human-computer interaction, it also introduces serious risks to security and public trust. In particular, the rise of audio deepfakes contributes to a form of \textit{impostor bias}, where individuals increasingly question the authenticity of audio content, undermining trust in legitimate recordings across domains such as journalism, legal evidence, and personal communication~\cite{CASU2024301795}. Recent work has shown that deepfakes are especially dangerous in subtle scenarios where background noise or speech hesitations may aid deception~\cite{borzi2024silence}.

As a result, robust audio deepfake detection has become a pressing need. Researchers have explored diverse strategies, including forensic analysis~\cite{guarnera2022mixture}, chatbot moderation~\cite{mentalhealth}, and the detection of synthetic media more broadly~\cite{guarnera2024mastering}. Other works have critically examined the state of voice spoofing research~\cite{Borzi_2022_CVPR}, highlighting open challenges in generalization and dataset bias.
Detecting audio deepfakes remains challenging. Many systems rely on handcrafted features (e.g., MFCCs, CQCCs) or spectrogram-based representations, which may fail to capture fine-grained temporal cues or generalize to unseen manipulations. Moreover, the detection task exhibits a core asymmetry: real audio spans a vast and diverse open domain (speakers, microphones, environments), whereas fake audio is generated by a finite, albeit expanding, set of models. This leads to frequent overfitting on known attacks and poor generalization to new ones.

In this work, we propose \textbf{RawNetLite}, a lightweight end-to-end architecture for audio deepfake detection, trained directly on raw waveforms. Unlike many prior approaches, RawNetLite does not rely on spectrogram transformations or large pretrained backbones. To improve generalization, we combine three strategies: (i) domain-mix training across multiple datasets (FakeOrReal, AVSpoof2021, and CodecFake), (ii) waveform-level data augmentation (pitch shift, noise injection, and time stretching), and (iii) Focal Loss to emphasize hard or ambiguous samples during training.

While the generalization gap in audio deepfake detection is well-known, our contribution lies in systematically exploring lightweight, reproducible solutions within a unified end-to-end framework. Rather than proposing entirely new components, we demonstrate how domain-mix training, Focal Loss, and waveform-level augmentation can be effectively combined to improve robustness under open-world conditions without relying on large pre-trained models or handcrafted features.

The main contributions are as follows:
\begin{itemize}
    \item We introduce \textbf{RawNetLite}, a compact and fully end-to-end architecture that processes raw audio waveforms without handcrafted features or spectrograms.
    \item We propose a robust training pipeline that integrates \textbf{domain-mix learning}, \textbf{Focal Loss}, and \textbf{waveform-level augmentation} to improve generalization under open-world conditions.
    \item We conduct extensive experiments across five test scenarios and three public datasets. Our model achieves over 99.7\% F1 and 0.25\% EER on in-domain data (FakeOrReal), and up to 83.4\% F1 and 16.4\% EER on challenging out-of-distribution test sets (AVSpoof2021 + CodecFake).
    \item \textbf{Differently from prior work}, we integrate all components into a unified, efficient, and deployable pipeline that balances generalization, performance, and computational efficiency—without relying on large self-supervised models.
\end{itemize}

The remainder of this paper is organized as follows: Section~\ref{sec:related} reviews related work in audio deepfake detection. Section~\ref{sec:data} describes the datasets and preprocessing strategy. Section~\ref{sec:method} details the RawNetLite architecture and training setup. Section~\ref{sec:experiments} presents our experimental results. Finally, Section~\ref{sec:conclusion} concludes the paper and outlines directions for future work.

\section{Related Work}
\label{sec:related}

The detection of audio deepfakes has received increasing attention due to the rapid advances in text-to-speech (TTS) and voice conversion (VC) technologies. Early systems relied on handcrafted features such as MFCCs, CQCCs, or prosodic cues~\cite{nagarsheth2017synthetic}, often processed by shallow classifiers or CNN-based backbones. While effective in constrained scenarios, these approaches typically underperform when exposed to spoofing conditions unseen during training~\cite{yi2023audio}.

To address these limitations, more recent work has shifted towards end-to-end deep learning approaches that operate directly on raw waveforms. RawNet and its variants~\cite{tak2021rawnet2, wang2023to} use convolutional and recurrent layers to extract spectral and temporal features from raw audio. Improvements have included orthogonal regularization~\cite{tak2021rawnet2}, SE-attention~\cite{lai2023attentive}, and advanced pooling strategies, but these architectures can be computationally heavy or prone to overfitting on seen attacks.

Hybrid systems like SpecRNet~\cite{janicki2022specrnet} combine spectrogram-based inputs with deep convolutional encoders. Self-supervised learning (SSL) models have also been explored: Tak et al.~\cite{tak2023asvspoof} used wav2vec 2.0 embeddings and ensemble fusion; Guo et al.~\cite{guo2024wavlm} proposed a WavLM-based system with attentive pooling; Zhang et al.~\cite{zhang2024sls} employed multi-layer selection over XLS-R transformers, achieving 1.92\% EER on ASVspoof2021 DF. These SSL systems deliver strong performance but rely on large pretrained models and considerable computational overhead.

Orthogonal to supervised pipelines, anomaly detection has also shown promise. Coletta et al.~\cite{coletta2025fpm} proposed a one-class student-teacher framework trained only on bonafide samples, reaching 9.1\% EER on FakeOrReal. Similarly, Kim et al.~\cite{kim2024oneclass} introduced an adaptive centroid shift strategy that attained 2.19\% EER on ASVspoof2021 DF using only real data.

Despite progress, a key challenge remains: generalizing to unseen spoofing techniques and acoustic conditions. Many systems achieve high in-domain accuracy but degrade in open-world evaluation. To mitigate this, recent strategies include domain adaptation, loss reweighting (e.g., Focal Loss~\cite{chettri2021end}), and audio augmentation (e.g., codec simulation, noise injection, time-stretching). However, their combined use in a unified, lightweight framework is still rare.

In this work, we address this gap by proposing RawNetLite, an efficient model that jointly integrates domain-mix training, Focal Loss, and waveform-level audio augmentation, yielding competitive in-domain performance (0.25\% EER on FakeOrReal) and strong generalization (16.4\% EER on AVSpoof2021+CodecFake), without relying on self-supervised transformers or handcrafted features.

\section{Datasets and Preprocessing}
\label{sec:data}

We conduct our study using three publicly available datasets: \textbf{FakeOrReal}~\cite{fakeorreal}, \textbf{AVSpoof2021}~\cite{todisco2019asvspoof}, and \textbf{CodecFake}~\cite{xie2024codecfake}. These datasets offer complementary characteristics in terms of spoofing strategies, compression artifacts, and recording conditions. They were selected to evaluate both in-domain performance and generalization to unseen conditions in cross-dataset scenarios. All datasets were preprocessed uniformly, and no sample used in testing was ever included in training or validation.

\vspace{-1mm}
\subsection{FakeOrReal (FoR)}
FakeOrReal served as our primary source of clean, high-quality audio samples. It contains real speech recorded under controlled conditions and synthetic speech generated using state-of-the-art text-to-speech (TTS) and voice conversion (VC) models\cite{fakeorreal}. The dataset was used in all training configurations: baseline (BCE), domain-mix, triple-domain, and augmentation-based training.

We applied a stratified 80/10/10 split for training, validation, and testing, ensuring class balance and full separation between phases. No augmented or codec-corrupted versions of this dataset were included. The corresponding statistics are shown in Table~\ref{tab:for_stats}.

\begin{table}[ht]
\centering
\caption{FakeOrReal – Number of samples per class and usage phase.}
\begin{tabular}{lcccc}
\hline
\textbf{Phase} & \textbf{Real} & \textbf{Fake} & \textbf{Used} & \textbf{Total} \\\hline
Training       & 25600 & 25600 & All                & 51200 \\
Validation     & 3200  & 3200  & All                & 6400 \\
Test (in-domain) & 3200 & 3200  & All                & 6400 \\\hline
\textbf{Total} & 32000 & 32000 & —                  & 64000 \\\hline
\end{tabular}
\label{tab:for_stats}
\vspace{-1mm}
\end{table}

\vspace{-1mm}
\subsection{AVSpoof2021}
AVSpoof2021 introduces greater variability in both spoofing techniques (including replay attacks, TTS, and VC) and recording environments\cite{todisco2019asvspoof}. It was used to test the model’s generalization capabilities and to inject diversity during training in domain-mix and triple-domain configurations.

We included a balanced subset of 6400 real and 6400 fake samples during training. The remaining portion—completely disjoint from the training set—was used for cross-dataset testing to evaluate performance on unseen manipulations. These phases and corresponding counts are detailed in Table~\ref{tab:avspoof_stats}.

\begin{table}[ht]
\scriptsize
\centering
\caption{AVSpoof2021 – Number of samples per class and usage phase.}
\begin{tabular}{lcccc}
\hline
\textbf{Phase} & \textbf{Real} & \textbf{Fake} & \textbf{Used in} & \textbf{Total} \\\hline
Training (Cross/Triple) & 6400  & 6400  & All excl. FOR & 12800 \\
Test (Cross-dataset)    & 22616 & 25000 & All excl. FOR & 47616 \\\hline
\textbf{Total}          & 29016 & 31400 & —             & 60416 \\\hline
\end{tabular}
\label{tab:avspoof_stats}
\vspace{-1mm}
\end{table}

\vspace{-1mm}
\subsection{CodecFake}
The CodecFake dataset was chosen to simulate real-world audio degradation scenarios caused by codec compression and transmission artifacts\cite{xie2024codecfake}. It contains both real and fake speech processed through various codec pipelines, introducing distortions not present in other datasets.

CodecFake was initially used exclusively for cross-dataset testing. In later configurations (triple-dataset and augmentation), a subset was included in training to improve robustness under degraded conditions. Balanced test sets were created from the remaining samples, as shown in Table~\ref{tab:codecfake_stats}.

\begin{table}[ht]
\scriptsize
\centering
\caption{CodecFake – Number of samples per class and usage phase.}
\begin{tabular}{lcccc}
\hline
\textbf{Phase} & \textbf{Real} & \textbf{Fake} & \textbf{Used} & \textbf{Total} \\\hline
Training (Triple/Augmented) & 6400  & 6400  & Triple, Augmented & 12800 \\
Test (Cross-dataset)        & 52000 & 50000 & All                & 102000 \\\hline
\textbf{Total}              & 58400 & 56400 & —                  & 114800 \\\hline
\end{tabular}
\label{tab:codecfake_stats}
\vspace{-1mm}
\end{table}

\vspace{-1mm}
\subsection{Preprocessing}
All audio samples were converted to mono and resampled to a fixed sampling rate of 16~kHz. This rate is widely adopted in the speech processing literature as a compromise between audio fidelity and computational efficiency, and it is used in several benchmark datasets for spoofing detection, including ASVspoof~\cite{todisco2019asvspoof} and FakeOrReal~\cite{fakeorreal}. Mono conversion ensures consistency across recordings and reduces redundant channel information, which is typically unnecessary for speaker-based deepfake detection.

Waveforms were normalized by dividing by their maximum absolute amplitude. This standard technique prevents amplitude differences from influencing the model and ensures numerical stability during training. Each waveform was trimmed or zero-padded to a fixed duration of exactly 3 seconds (48000 samples), in line with common practices in end-to-end audio processing~\cite{wang2023to}. This fixed-length representation allows for efficient mini-batch processing and ensures compatibility with convolutional and recurrent layers.

We avoided handcrafted preprocessing steps such as spectrograms or MFCCs, opting instead to feed raw audio directly into the model. This design choice is motivated by previous work demonstrating that end-to-end models can automatically learn discriminative features from raw signals, often outperforming traditional pipelines in open-domain tasks~\cite{lai2023attentive}.

\vspace{-1mm}
\subsection{Data Augmentation}
In the augmentation-based training setup, we introduced synthetic variation into the training data to improve generalization under real-world acoustic conditions. Augmentations were applied on-the-fly during training using the \texttt{audiomentations} library. Each transformation was applied with a probability of $p = 0.5$:
\begin{itemize}
    \item \textbf{Pitch shift}: random semitone shift in the range $[-2, +2]$ to simulate speaker pitch variation and emotional tone.
    \item \textbf{Time stretch}: random tempo changes drawn from a uniform distribution over $[0.9, 1.1]$, simulating variable speaking rates or playback distortion.
    \item \textbf{Gaussian noise}: low-amplitude additive noise with amplitude $\in [0.001, 0.015]$, emulating environmental interference or low-quality transmission.
\end{itemize}

These transformations reflect real-world distortions encountered in telephony, VoIP communications, and online meetings, where spoofed speech may appear under degraded conditions. Similar augmentation strategies have proven effective in prior work on robust audio classification and spoof detection~\cite{zhang2021deep, yi2023audio}.

\vspace{-1mm}
\subsection{Separation of Training and Testing Data}
In all configurations, we ensured strict disjointness between training, validation, and test sets. Cross-dataset evaluations were conducted exclusively on data from datasets not used in training (e.g., AVSpoof2021 or CodecFake), and in-domain test splits (FakeOrReal) were held out from training and validation. This design guarantees that performance metrics genuinely reflect out-of-distribution generalization capabilities.

Details on the dataset splits and sample distributions are presented in Tables~\ref{tab:for_stats}--\ref{tab:codecfake_stats}, while the impact of preprocessing and augmentation is discussed in Section~\ref{sec:experiments}.

\section{Proposed Method}
\label{sec:method}

We propose a lightweight end-to-end architecture for audio deepfake detection based on raw waveforms, inspired by RawNet~\cite{wang2023to}. Our model, termed \textit{RawNetLite}, is designed for binary classification and tailored for efficient deployment in real-world scenarios. It directly learns from raw audio, avoiding the need for handcrafted feature extraction or spectral transformations.

The model takes as input a mono audio waveform of fixed duration, sampled at 16~kHz and normalized to the range $[-1, 1]$. Each input has shape $[1 \times 48000]$, corresponding to 3 seconds of audio.

\vspace{-1mm}
\subsection{Architecture Overview}

As shown in Figure \ref{fig:rawnetlite} architecture starts with a 1D convolutional layer (64 filters, kernel size 3, stride 1, padding 1), followed by three residual blocks, each containing two convolutional layers with identical settings and ReLU activations. These layers extract hierarchical local features directly from the raw waveform. The output is reduced via adaptive average pooling, then passed to a bidirectional GRU with 128 units per direction to capture long-range speaker and prosody dynamics. Finally, two fully connected layers produce a scalar probability via a sigmoid activation. This design balances performance and efficiency: residual blocks model local patterns, while the GRU captures temporal dependencies. Compared to larger architectures (e.g., RawNet2, transformer-based models), RawNetLite offers competitive accuracy with lower computational cost, making it suitable for real-time or embedded applications.

\usetikzlibrary{positioning, arrows.meta, shapes}
\begin{figure}[ht]
\centering
\begin{tikzpicture}[
    node distance=0.3cm,
    every node/.style={align=center, minimum width=6.2cm, minimum height=0.4cm, font=\small},
    arrow/.style={-{Latex[length=2mm]}, thick},
    input/.style={draw, fill=cyan!20, rounded corners},
    conv/.style={draw, fill=green!20, rounded corners},
    res/.style={draw, fill=green!10, rounded corners},
    pool/.style={draw, fill=cyan!10, rounded corners},
    gru/.style={draw, fill=blue!15, rounded corners},
    fc/.style={draw, fill=blue!5, rounded corners},
    output/.style={draw, fill=gray!20, rounded corners}
]

\node[input] (input) {Raw Audio - $[1 \times 48000]$};
\node[conv, below=of input] (conv) {Conv1D + BN + ReLU};
\node[res, below=of conv] (res1) {Residual Block 1};
\node[res, below=of res1] (res2) {Residual Block 2};
\node[res, below=of res2] (res3) {Residual Block 3};
\node[pool, below=of res3] (pool) {Adaptive AvgPool1D};
\node[gru, below=of pool] (gru) {Bidirectional GRU - $128 \times 2$ units};
\node[fc, below=of gru] (fc) {FC + ReLU + Sigmoid};
\node[output, below=of fc] (out) {Output Probability};

\draw[arrow] (input) -- (conv);
\draw[arrow] (conv) -- (res1);
\draw[arrow] (res1) -- (res2);
\draw[arrow] (res2) -- (res3);
\draw[arrow] (res3) -- (pool);
\draw[arrow] (pool) -- (gru);
\draw[arrow] (gru) -- (fc);
\draw[arrow] (fc) -- (out);

\end{tikzpicture}
\caption{Architecture of the proposed \textit{RawNetLite} model for audio deepfake detection.}
\label{fig:rawnetlite}
\vspace{-1mm}
\end{figure}
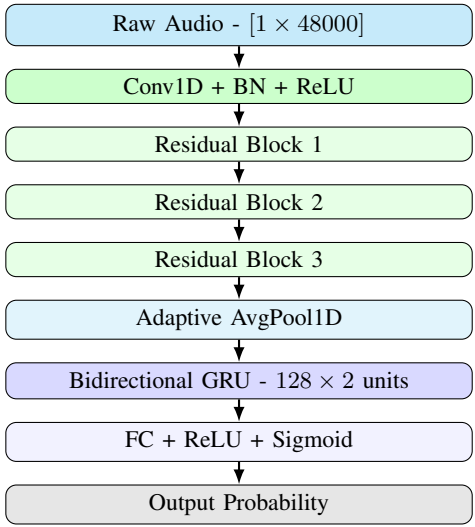

\vspace{-1mm}
\subsection{Training Details}
Training was conducted on a workstation equipped with an NVIDIA RTX A6000 GPU (48GB VRAM) and a 32-core AMD Ryzen Threadripper PRO 3975WX CPU.
The model is trained using the Adam optimizer with a learning rate of $10^{-4}$ and a batch size of 16. Unless otherwise specified, training proceeds for 10 epochs, with early stopping based on validation F1-score.
In the augmented training setting, we introduce randomly perturbed versions of the original waveforms using pitch shifting, time stretching, and additive Gaussian noise. These augmentations are applied on-the-fly and mixed with unmodified samples, preserving diversity while enriching the training distribution.

\vspace{-1mm}
\subsection{Binary Cross-Entropy Loss}
As a baseline, we employ the standard Binary Cross-Entropy (BCE) loss:
\begin{formula}
\mathcal{L}_{\text{BCE}} = -\left[ y \cdot \log(p) + (1 - y) \cdot \log(1 - p) \right]
\end{formula}
where \( y \in \{0,1\} \) is the true label and \( p \in (0,1) \) is the predicted probability output by the model. BCE treats all errors equally, which may be suboptimal in settings with class imbalance or distributional shifts.

\vspace{-1mm}
\subsection{Focal Loss}
To focus training on difficult or rare examples, particularly under cross-dataset conditions, we adopt the Focal Loss~\cite{lin2017focal}. This loss down-weights easy examples and amplifies the impact of misclassified or ambiguous samples:
\begin{formula}
\mathcal{L}_{\text{Focal}} = -\alpha_t (1 - p_t)^{\gamma} \log(p_t)
\end{formula}
where:
\begin{itemize}
  \item \( p_t = p \) if \( y = 1 \), and \( p_t = 1 - p \) if \( y = 0 \),
  \item \( \gamma = 2 \) is the focusing parameter,
  \item \( \alpha_t = 0.25 \) balances the class weights.
\end{itemize}

Focal Loss proved especially effective when used in combination with domain-mix training and augmentation, improving model sensitivity to spoofed samples from unseen distributions.

\vspace{-1mm}
\subsection{Evaluation Metrics}
We evaluate the model using Accuracy, Precision, Recall, and F1-score~\cite{sokolova2009systematic}, and Equal Error Rate (EER)~\cite{martin1997det}.

These metrics allow us to assess the model's performance both in detecting fake samples (recall and precision) and its overall classification quality (F1-score, accuracy). In addition, the Equal Error Rate (EER) provides a threshold-independent measure commonly used in biometric verification and spoof detection. It corresponds to the operating point at which the rate of false positives equals the rate of false negatives, offering a compact summary of a system’s trade-off between sensitivity and specificity.

\smallskip
This architecture, coupled with an appropriate loss function and diverse training inputs, enables the model to learn robust features directly from raw audio. As demonstrated in Section~\ref{sec:experiments}, this approach enhances generalization and maintains strong performance under challenging open-world scenarios.

\section{Experiments and Results}
\label{sec:experiments}

\vspace{-1mm}
\subsection{Training Setup}
We trained the \textit{RawNetLite} model using combinations of the FakeOrReal, AVSpoof2021, and CodecFake datasets. All datasets were preprocessed uniformly and balanced to include the same number of real and fake samples. The optimizer was Adam with a learning rate of $10^{-4}$ and a batch size of 16. We used early stopping based on validation F1-score and applied Focal Loss for all experiments. All experiments were run for a maximum of 10 epochs, and we found that the best results were obtained without requiring longer training. To improve generalization, we adopted a \textit{domain-mix training} strategy, including a subset of AVSpoof2021 or CodecFake alongside FakeOrReal. Test splits remained strictly disjoint, ensuring no overlap in generation methods or identities.

\vspace{-1mm}
\subsection{Evaluation Protocol}
To evaluate performance, we define three experimental settings:

\begin{itemize}
\item \textbf{In-domain}: training and testing are conducted on the same dataset (FakeOrReal), with disjoint splits.
\item \textbf{Cross-domain}: training is performed on a subset of two datasets, and testing is done on a third dataset or a combination involving an unseen domain (e.g., training on FakeOrReal + AVSpoof2021 and testing on CodecFake or AVSpoof2021 + CodecFake).
\item \textbf{Triple-domain}: training and testing involve all three datasets. This configuration maximizes diversity and is used to simulate realistic deployment conditions.
\end{itemize}

\vspace{-1mm}
\subsection{In-domain Performance}
On the internal test split of the FakeOrReal dataset, as shown in the Table \ref{tab:for_rawnetlite}, the model achieved an F1-score of 99.2\% and an overall accuracy above 99.2\%. These results indicate that the model is effective at distinguishing between real and fake speech when the data distribution matches the training set.

\begin{table}[ht]
\centering
\caption{Results on FakeOrReal (RawNetLite - Baseline)}
\begin{tabular}{lcccc}
\hline
\textbf{Class} & \textbf{Precision} & \textbf{Recall} & \textbf{F1-score} & \textbf{Support} \\\hline
Real           & 0.9998             & 0.9856          & 0.9926            & 32496 \\
Fake           & 0.9858             & 0.9998          & 0.9927            & 32428 \\\hline
\textbf{Accuracy} & \multicolumn{4}{c}{0.9927} \\
\textbf{EER}      & \multicolumn{4}{c}{0.0029} \\\hline
\end{tabular}
\label{tab:for_rawnetlite}
\vspace{-1mm}
\end{table}

\subsection{Cross-domain Evaluation}

\subsubsection{CodecFake}
To evaluate the generalization capacity of the model, we tested it on a balanced subset of the CodecFake dataset, using 52000 real and 50000 fake samples. Table~\ref{tab:codecfake_rawnetlite} shows that the baseline model performs poorly on this dataset, achieving a fake F1-score of only 17.8\% and an EER of 50.2\%. To assess whether domain-mixing can help, we also trained a version of the model using both FakeOrReal and AVSpoof2021. As shown in Table~\ref{tab:codecfake_results}, performance remains low, with minimal gains in fake recall and F1-score (17.9\%) and a similarly high EER (50.4\%).

These results highlight the difficulty of generalizing to heavily compressed or codec-manipulated speech without direct exposure during training, even when using diverse data and Focal Loss.

\begin{table}[ht]
\centering
\caption{Results on CodecFake (RawNetLite - Baseline)}
\begin{tabular}{lcccc}
\hline
\textbf{Class} & \textbf{Precision} & \textbf{Recall} & \textbf{F1-score} & \textbf{Support} \\\hline
Real           & 0.4922             & 0.8573          & 0.6253            & 52206 \\
Fake           & 0.4385             & 0.1119          & 0.1783            & 52000 \\\hline
\textbf{Accuracy} & \multicolumn{4}{c}{0.4853} \\
\textbf{EER}      & \multicolumn{4}{c}{0.5019} \\\hline
\end{tabular}
\label{tab:codecfake_rawnetlite}
\vspace{-1mm}
\end{table}

\begin{table}[ht]
\centering
\caption{Results on CodecFake (RawNetLite - Cross-domain training)}
\begin{tabular}{lcccc}
\hline
\textbf{Class} & \textbf{Precision} & \textbf{Recall} & \textbf{F1-score} & \textbf{Support} \\\hline
Real           & 0.4744             & 0.7921          & 0.5934            & 52206 \\
Fake           & 0.3632             & 0.1190          & 0.1793            & 52000 \\\hline
\textbf{Accuracy} & \multicolumn{4}{c}{0.4562} \\
\textbf{EER}      & \multicolumn{4}{c}{0.5039} \\\hline
\end{tabular}
\label{tab:codecfake_results}
\vspace{-1mm}
\end{table}

\subsubsection{AVSpoof2021}
We further evaluated the model on the AVSpoof2021 dataset, using 22617 real and 25000 fake samples (balanced). As shown in Tables~\ref{tab:avspoof_results} and ~\ref{tab:avspoof_results_cross}, the baseline model trained only on FakeOrReal fails to generalize to AVSpoof2021, achieving 55.7\% accuracy, a fake F1-score of 34.4\%, and an EER of 33.2\%, with strong bias towards real speech (92.9\% recall). In contrast, the cross-domain training configuration substantially improves performance, raising the fake F1-score to 82.4\% and lowering the EER to 16.6\%, confirming the effectiveness of domain-mix strategies.

\begin{table}[ht]
\centering
\caption{Results on AVSpoof2021 (RawNetLite - baseline)}
\begin{tabular}{lcccc}
\hline
\textbf{Class} & \textbf{Precision} & \textbf{Recall} & \textbf{F1-score} & \textbf{Support} \\\hline
Real           & 0.5190             & 0.9288          & 0.6659            & 22616 \\
Fake           & 0.7745             & 0.2212          & 0.3441            & 25000 \\\hline
\textbf{Accuracy} & \multicolumn{4}{c}{0.5573} \\
\textbf{EER}      & \multicolumn{4}{c}{0.3322} \\\hline
\end{tabular}
\label{tab:avspoof_results}
\vspace{-1mm}
\end{table}

\begin{table}[ht]
\centering
\caption{Results on AVSpoof2021 (RawNetLite - Cross-domain training)}
\begin{tabular}{lcccc}
\hline
\textbf{Class} & \textbf{Precision} & \textbf{Recall} & \textbf{F1-score} & \textbf{Support} \\\hline
Real           & 0.7809             & 0.8821          & 0.8284            & 22616 \\
Fake           & 0.8792             & 0.7760          & 0.8244            & 25000 \\\hline
\textbf{Accuracy} & \multicolumn{4}{c}{0.8264} \\
\textbf{EER}      & \multicolumn{4}{c}{0.1660} \\\hline
\end{tabular}
\label{tab:avspoof_results_cross}
\vspace{-1mm}
\end{table}

As shown in Table~\ref{tab:domainmix_comparison}, incorporating 20\% of AVSpoof2021 into the training set led to a marked increase in detection performance on the same dataset. The most notable gain was in fake recall (+55.4 percentage points), demonstrating that domain mixing helps the model adapt to spoofing techniques unseen during initial training.

\begin{table}[ht]
\centering
\caption{Cross-dataset results on AVSpoof2021 before and after domain-mix training.}
\begin{tabular}{lcccc}
\hline
\textbf{Metric} & \textbf{Pre-mix} & \textbf{Post-mix} & \textbf{Change} \\\hline
Fake Precision   & 0.7745 & 0.8792 & \textbf{+0.1047} \\
Fake Recall      & 0.2212 & 0.7760 & \textbf{+0.5548} \\
Fake F1-score    & 0.3441 & 0.8244 & \textbf{+0.4803} \\
Accuracy         & 0.5573 & 0.8264 & \textbf{+0.2691} \\
Macro F1         & 0.5050 & 0.8264 & \textbf{+0.3214} \\
EER              & 0.3322 & 0.1660 & \textbf{--0.1662} \\\hline
\end{tabular}
\label{tab:domainmix_comparison}
\vspace{-1mm}
\end{table}

\subsection{Effect of Focal Loss}

To further improve detection of challenging spoofed audio, we replaced the standard binary cross-entropy loss with Focal Loss~\cite{lin2017focal}, using hyperparameters $\gamma = 2$ and $\alpha = 0.25$. This modification encourages the model to focus more on misclassified and difficult examples, which are especially common in cross-dataset scenarios. As shown in Table~\ref{tab:focalloss_effect}, applying Focal Loss resulted in a substantial performance gain on AVSpoof2021. The recall on spoofed samples increased from 44.9\% to 71.9\%, while the F1-score rose from 55.8\% to 79.5\%. Accuracy also improved significantly, reaching 80.6\% overall. These results indicate that Focal Loss helps the model better generalize to out-of-distribution manipulations and previously unseen generation methods.

Importantly, the use of Focal Loss did not degrade in-domain performance, with the model maintaining an F1-score above 96\% on the FakeOrReal test set. This demonstrates that hard-example mining via Focal Loss is a valuable strategy in training robust deepfake detection systems.

\begin{table}[ht]
\centering
\caption{Effect of Focal Loss on AVSpoof2021.}
\begin{tabular}{lccc}
\hline
\textbf{Metric} & \textbf{BCE + Mix} & \textbf{Focal + Mix} & \textbf{Change} \\\hline
Fake Precision  & 0.7369 & \textbf{0.8899} & +0.1530 \\
Fake Recall     & 0.4493 & \textbf{0.7189} & +0.2696 \\
Fake F1         & 0.5582 & \textbf{0.7953} & +0.2371 \\
Accuracy        & 0.6444 & \textbf{0.8057} & +0.1613 \\
Macro F1        & 0.6304 & \textbf{0.8052} & +0.1748 \\
EER             & 0.1660 & \textbf{0.1734} & +0.0074 \\\hline
\end{tabular}
\label{tab:focalloss_effect}
\vspace{-1mm}
\end{table}

Table~\ref{tab:cross_focal_results} summarizes the performance of our models across different test sets and training durations. All models were trained using Focal Loss on a combination of FakeOrReal and AVSpoof2021.

The model shows excellent in-domain performance on FakeOrReal, achieving over 99.8\% F1-score. On AVSpoof2021, training with Focal Loss substantially improves recall and precision, reaching an F1-score of 79.5\%. However, results on the cross-dataset test set reveal a trade-off: while the model performs well on known domains, it exhibits reduced performance on unseen domains like CodecFake. Specifically, fake recall on CodecFake remains limited (9.56\%), indicating a form of domain-specific overfitting.

These results highlight the importance of balancing training depth with exposure to diverse spoofing domains to preserve generalization under open-world conditions.

\begin{table}[ht]
\centering
\caption{Performance on various test sets using Focal Loss (Cross-domain training).}
\begin{tabular}{lccccc}
\hline
\textbf{Test Set} & \textbf{P-fake} & \textbf{R-fake} & \textbf{F1-fake} & \textbf{Accuracy} & \textbf{EER} \\\hline
FakeOrReal         & 99.71\% & 99.79\% & 99.75\% & 99.75\% & 0.25\% \\
AVSpoof2021        & 88.99\% & 71.89\% & 79.53\% & 80.57\% & 17.34\% \\
CodecFake          & 35.99\% & 9.56\%  & 15.11\% & 46.38\% & 50.68\% \\
Cross-domain    & 75.13\% & 40.80\% & 52.89\% & 63.65\% & 36.88\% \\\hline
\end{tabular}
\label{tab:cross_focal_results}
\vspace{-1mm}
\end{table}

\subsection{Triple-domain Training}

To improve cross-dataset generalization, we trained the model on a combined dataset comprising FakeOrReal, AVSpoof2021, and CodecFake. This configuration significantly improved robustness on previously unseen spoofing conditions. In this setting, the model achieved strong performance on CodecFake, with an F1-score above 78\%, indicating improved robustness to codec-induced distortions.

As shown in Table~\ref{tab:triple_focal_results}, this setup yielded a fake F1-score of 78.6\% on CodecFake—compared to 15.1\% when CodecFake was excluded. It also improved generalization on a combined test set (AVSpoof + CodecFake), reaching an overall accuracy of 83.9\% and a fake F1-score of 83.4\%.

\begin{table}[ht]
\centering
\caption{Performance on various test sets using Focal Loss (Triple-domain training).}
\begin{tabular}{lccccc}
\hline
\textbf{Test Set} & \textbf{P-fake} & \textbf{R-fake} & \textbf{F1-fake} & \textbf{Accuracy} & \textbf{EER} \\\hline
FakeOrReal         & 98.80\% & 95.33\% & 97.03\% & 97.08\% & 2.66\% \\
AVSpoof2021        & 92.92\% & 51.10\% & 65.94\% & 72.28\% & 19.96\% \\
CodecFake          & 75.72\% & 81.57\% & 78.53\% & 77.75\% & 21.13\% \\
Cross-domain    & 81.38\% & 66.49\% & 73.18\% & 75.64\% & 25.35\% \\
Triple-domain   & 85.81\% & 80.34\% & 83.40\% & 83.85\% & 16.44\% \\\hline
\end{tabular}
\label{tab:triple_focal_results}
\vspace{-1mm}
\end{table}

\subsection{Effect of Audio Augmentation}

To further enhance robustness under real-world variability, we trained the model using both original and augmented audio samples. The training set combined FakeOrReal and AVSpoof2021, and each waveform was duplicated and subjected to a random augmentation pipeline during training. This effectively doubled the number of training samples, while maintaining class balance.

Augmentations were applied using the \texttt{audiomentations} library, with three stochastic transformations:

\begin{itemize}
    \item \textbf{Pitch Shift:} randomly shifts the pitch of the waveform within a range of $[-2, +2]$ semitones.
    \item \textbf{Time Stretch:} simulates temporal variation by stretching the waveform with a factor randomly chosen from $[0.9, 1.1]$.
    \item \textbf{Gaussian Noise:} adds low-amplitude noise with an amplitude sampled from $[0.001, 0.015]$.
\end{itemize}

Each transformation was applied with probability $p=0.5$ to each input. The combination of pitch variation, local timing distortion, and low-level background noise aims to simulate acoustic conditions encountered in real-world recordings or during transmission (e.g., online meetings, mobile calls).

As shown in Table~\ref{tab:aug_results}, this setup resulted in a fake F1-score of 78.05\% on the test set and a balanced accuracy of 77.74\%. These results demonstrate that audio augmentation improves generalization without requiring new datasets, and helps the model become more robust to subtle distortions or unseen acoustic environments.

\begin{table}[ht]
\centering
\caption{Results using audio augmentation on Triple-dataset}
\begin{tabular}{lcccc}
\hline
\textbf{Class} & \textbf{Precision} & \textbf{Recall} & \textbf{F1-score} & \textbf{Support} \\\hline
Real       & 0.8058             & 0.8158          & 0.8108            & 25000 \\
Fake       & 0.8135             & 0.8034          & 0.8084            & 25000 \\\hline
\textbf{Accuracy} & \multicolumn{4}{c}{0.8096} \\
\textbf{EER}      & \multicolumn{4}{c}{0.1913} \\\hline
\end{tabular}
\label{tab:aug_results}
\vspace{-1mm}
\end{table}

We also explored the impact of waveform-level audio augmentation as a lightweight alternative to domain-mix and multi-dataset training. Using a combination of pitch shifting, time stretching, and Gaussian noise, applied on-the-fly to the training data, we observed consistent improvements in cross-dataset generalization.

As shown in Table~\ref{tab:aug_training_results}, the model achieved a fake F1-score of 74.66\% on CodecFake (vs. 15.1\% without augmentation) and 74.25\% on the combined FOR+AVSpoof2021 test set. Most notably, the performance on the triple-dataset test set exceeded 80\% F1, surpassing even the triple-domain training baseline. These results suggest that simple data augmentation can effectively simulate real-world variability and improve model robustness, without requiring additional datasets or architectural changes.

\begin{table}[ht]
\centering
\caption{Performance of the model trained with audio augmentation (FakeOrReal + AVSpoof2021, 10 epochs).}
\begin{tabular}{lcccc}
\hline
\textbf{Test Set} & \textbf{P-fake} & \textbf{R-fake} & \textbf{F1-fake} & \textbf{Accuracy} \\\hline
FakeOrReal         & 96.11\% & 98.79\% & 97.43\% & 97.40\% \\
CodecFake          & 70.22\% & 79.71\% & 74.66\% & 73.00\% \\
AVSpoof2021        & 91.84\% & 61.98\% & 74.01\% & 77.15\% \\
Cross (FOR+AVS)    & 78.30\% & 70.59\% & 74.25\% & 75.51\% \\
Triple (FOR+AVS+Codec) & 81.35\% & 80.34\% & 80.84\% & 80.96\% \\\hline
\end{tabular}
\label{tab:aug_training_results}
\vspace{-1mm}
\end{table}

\begin{figure}[ht]
    \centering
    \includegraphics[width=\linewidth]{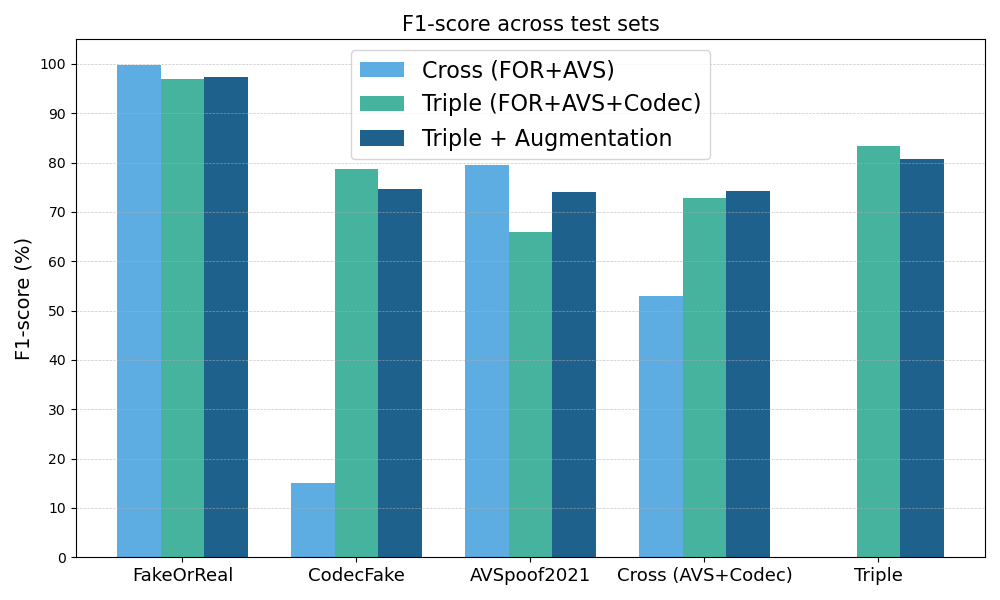}
    \caption{Comparison of fake F1-scores across test sets for models trained with Cross and Triple-domain strategies.}
    \label{fig:f1_comparison_augmented}
\end{figure}
\begin{figure}[ht]
    \centering
    \includegraphics[width=\linewidth]{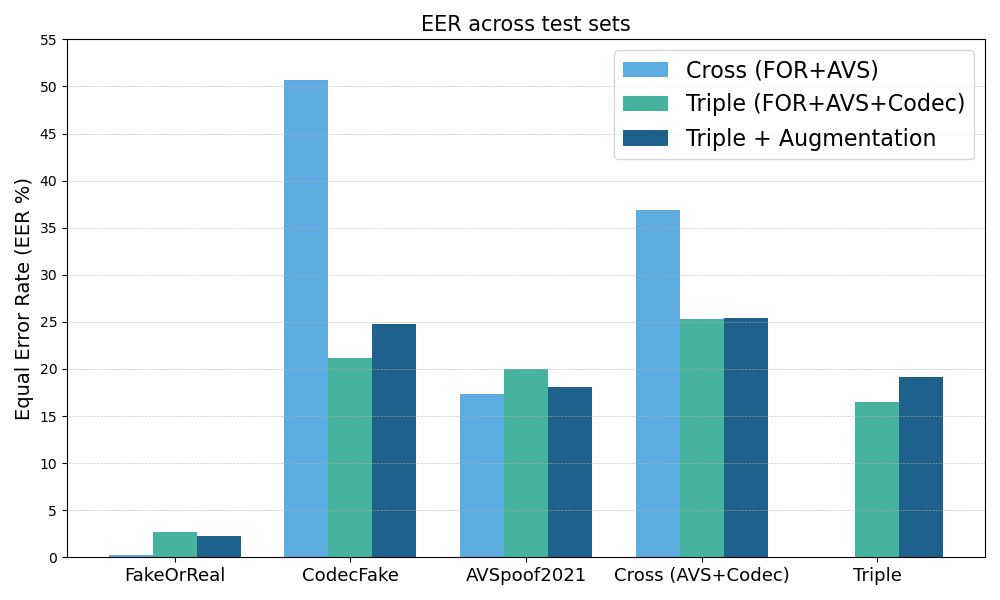}
    \caption{Comparison of EER across test sets for models trained with Cross and Triple-domain strategies.}
    \label{fig:eer_comparison_augmented}
    \vspace{-1mm}
\end{figure}

Figures~\ref{fig:f1_comparison_augmented} and \ref{fig:eer_comparison_augmented} summarizes the fake F1-scores and the EER scores achieved across five test sets for models trained under the following configurations:
\begin{itemize}
    \item \textbf{Cross-domain training:} using FakeOrReal and AVSpoof2021;
    \item \textbf{Triple-domain training:} using FakeOrReal, AVSpoof2021, and CodecFake;
    \item \textbf{Triple-domain with augmentation:} same as above, with waveform-level audio augmentations applied during training.
\end{itemize}

The figures highlights several key trends. While the model achieves strong in-domain performance on FakeOrReal, this does not always translate into high cross-dataset generalization. Triple-domain training significantly improves performance on CodecFake, and the addition of waveform-level audio augmentation further enhances robustness across all test settings. Notably, the triple augmented model achieves over 80\% F1-score on the joint AVSpoof + CodecFake test set, marking the highest generalization performance in our study. This confirms the effectiveness of combining diverse training data with realistic augmentations to simulate open-world variability.

\subsection{Discussion}
These results highlight the core challenge of audio deepfake detection: the domain gap between training and testing distributions. While our model achieves near-perfect in-domain performance (99.2\% F1, 0.29\% EER on FakeOrReal), cross-dataset results reveal significant generalization issues, with F1-scores of 34.4\% (AVSpoof2021) and 17.8\% (CodecFake). This reflects a typical open-set scenario: real audio is highly diverse, whereas fake audio is generated by a finite set of models, causing overfitting risks. To mitigate this, we introduced domain-mix training, improving fake recall on AVSpoof2021 by over 55 points, and adopted Focal Loss, boosting the fake F1 from 55.8\% to 79.5\%. Expanding to a triple-domain setup further improved generalization, reaching 83.4\% F1 on the combined AVSpoof + CodecFake test set (EER 16.4\%).
Waveform-level audio augmentation (pitch shift, time stretch, noise) proved equally effective, surpassing 80\% fake F1 on the triple test set and improving robustness on CodecFake (74.7\% fake F1, EER 19.1\%). Tables~\ref{tab:eer_for_comparison} and~\ref{tab:eer_avspoof_comparison} show that RawNetLite achieves competitive performance against recent state-of-the-art methods: 0.29\% EER on FakeOrReal (baseline), 2.25\% (triple+augmentation), and 16.6\%--18.1\% EER on AVSpoof2021. These results confirm that combining lightweight models with data diversity and augmentation yields a strong robustness-efficiency trade-off.

Future work will explore domain-adversarial training, self-supervised pretraining, and explicit open-set recognition frameworks.

\begin{table}[ht]
\centering
\caption{Comparison of Equal Error Rate (EER) on the FakeOrReal (FoR) dataset.}
\resizebox{\linewidth}{!}{%
\begin{tabular}{lcc}
\hline
\textbf{Method} & \textbf{Architecture / Strategy} & \textbf{EER (\%)} \\\hline
RawNet2 Baseline~\cite{tak2021rawnet2}         & Raw waveform + GRU + Attention        & 25.1 \\
FPM Anomaly Detector~\cite{coletta2025fpm}     & Feature Pyramid Matching (1-class)    & 9.1  \\
MoE LCNN Fusion~\cite{negroni2024moe}          & MoE (enhanced) + Gating Net  & 2.74 \\
\textbf{Ours (RawNetLite)}                     & Raw waveform + GRU (no mix)           & \textbf{0.29} \\
\textbf{Ours (Triple + Augmented)}             & + Mix + Focal + Augmentations         & \textbf{2.25} \\\hline
\end{tabular}
}
\label{tab:eer_for_comparison}
\vspace{-1mm}
\end{table}

\begin{table}[ht]
\centering
\caption{Comparison of Equal Error Rate (EER) on the AVSpoof2021 dataset.}
\resizebox{\linewidth}{!}{%
\begin{tabular}{lcc}
\hline
\textbf{Method} & \textbf{Architecture / Strategy} & \textbf{EER (\%)} \\\hline
ACS One-Class~\cite{kim2024oneclass}            & Adaptive Centroid Shift (1-class)    & 2.19 \\
XLS-R + SLS~\cite{zhang2024sls}                 & Self-Sup. XLS-R + Layer Selection    & 1.92 \\
WavLM + MFA~\cite{guo2024wavlm}                 & WavLM + Multi-Fusion Attention       & 2.56 \\
Tak et al.~\cite{tak2021rawnet2}                & XLS-R + RawNet2 + AASIST Backend     & 2.85 \\
\textbf{Ours (RawNetLite Cross-domain)}                    & Raw waveform + GRU + AVS mix                            & \textbf{16.6} \\
\textbf{Ours (Cross + Focal)}                   & + Focal Loss                         & \textbf{17.3} \\
\textbf{Ours (Triple + Focal)}                  & + CodecFake                          & \textbf{20.0} \\
\textbf{Ours (Triple + Augmented)}              & + Augmentations                      & \textbf{18.1} \\\hline
\end{tabular}
}
\label{tab:eer_avspoof_comparison}
\vspace{-1mm}
\end{table}

\section{Conclusion and Future Work}
\label{sec:conclusion}

In this work, we proposed \textit{RawNetLite}, a lightweight end-to-end model for audio deepfake detection operating directly on raw waveforms, achieving over 99\% F1-score on in-domain data (FakeOrReal) with minimal preprocessing.

Our experiments highlighted the fundamental challenges of open-world detection: while strong performance was achieved on seen data, generalization to unseen attacks (AVSpoof2021, CodecFake) remained difficult. To address this, we introduced a combined strategy of domain-mix training, Focal Loss, and waveform-level augmentation. This approach improved robustness significantly, reaching up to 83.4\% F1 on the AVSpoof + CodecFake set, and demonstrated that simple augmentations can boost cross-domain performance beyond 80\% without additional labeled data.

These findings emphasize the value of diverse training and tailored objective functions. As future work, we plan to explore domain-adversarial learning, self-supervised pretraining, open-set recognition, and augmentations simulating room acoustics and transmission artifacts, to further bridge the gap toward real-world deployment.

\balance{
\bibliographystyle{IEEEtran}
\bibliography{main}

\begin{thebibliography}{10}
\providecommand{\url}[1]{#1}
\csname url@samestyle\endcsname
\providecommand{\newblock}{\relax}
\providecommand{\bibinfo}[2]{#2}
\providecommand{\BIBentrySTDinterwordspacing}{\spaceskip=0pt\relax}
\providecommand{\BIBentryALTinterwordstretchfactor}{4}
\providecommand{\BIBentryALTinterwordspacing}{\spaceskip=\fontdimen2\font plus
\BIBentryALTinterwordstretchfactor\fontdimen3\font minus \fontdimen4\font\relax}
\providecommand{\BIBforeignlanguage}[2]{{%
\expandafter\ifx\csname l@#1\endcsname\relax
\typeout{** WARNING: IEEEtran.bst: No hyphenation pattern has been}%
\typeout{** loaded for the language `#1'. Using the pattern for}%
\typeout{** the default language instead.}%
\else
\language=\csname l@#1\endcsname
\fi
#2}}
\providecommand{\BIBdecl}{\relax}
\BIBdecl

\bibitem{amerini2025deepfake}
I.~Amerini, M.~Barni, S.~Battiato, P.~Bestagini, G.~Boato, V.~Bruni, R.~Caldelli, F.~De~Natale, R.~De~Nicola, L.~Guarnera \emph{et~al.}, ``Deepfake media forensics: Status and future challenges,'' \emph{Journal of Imaging}, vol.~11, no.~3, p.~73, 2025.

\bibitem{tolosana2020deepfakes}
R.~Tolosana, R.~Vera-Rodriguez, J.~Fierrez, A.~Morales, and J.~Ortega-Garcia, ``Deepfakes and beyond: A survey of face manipulation and fake detection,'' \emph{Information Fusion}, vol.~64, pp. 131--148, 2020.

\bibitem{mirsky2021creation}
Y.~Mirsky and W.~Lee, ``The creation and detection of deepfakes: A survey,'' \emph{ACM Computing Surveys (CSUR)}, vol.~54, no.~1, pp. 1--41, 2021.

\bibitem{yi2023audio}
J.~Yi, C.~Wang, J.~Tao, X.~Zhang, C.~Y. Zhang, and Y.~Zhao, ``Audio deepfake detection: A survey,'' \emph{arXiv preprint arXiv:2308.14970}, 2023.

\bibitem{CASU2024301795}
M.~Casu, L.~Guarnera, P.~Caponnetto, and S.~Battiato, ``{GenAI mirage: The impostor bias and the deepfake detection challenge in the era of artificial illusions},'' \emph{Forensic Science International: Digital Investigation}, vol.~50, p. 301795, 2024.

\bibitem{borzi2024silence}
S.~Borz{\`i}, L.~Mongelli, F.~Stanco, S.~Battiato, and D.~Allegra, ``Breaking the silence: Detecting ai-converted voices in the quietest moments,'' in \emph{Pattern Recognition. ICPR 2024 International Workshops and Challenges}, S.~Palaiahnakote, S.~Schuckers, J.-M. Ogier, P.~Bhattacharya, U.~Pal, and S.~Bhattacharya, Eds.\hskip 1em plus 0.5em minus 0.4em\relax Cham: Springer Nature Switzerland, 2025, pp. 185--194.

\bibitem{guarnera2022mixture}
L.~Guarnera, O.~Giudice, and S.~Battiato, ``Deepfake style transfer mixture: A first forensic ballistics study on synthetic images,'' in \emph{Image Analysis and Processing -- ICIAP 2022}, vol. 13232.\hskip 1em plus 0.5em minus 0.4em\relax Cham: Springer International Publishing, 2022, pp. 151--163.

\bibitem{mentalhealth}
M.~Casu, S.~Triscari, S.~Battiato, L.~Guarnera, and P.~Caponnetto, ``{AI Chatbots for Mental Health: A Scoping Review of Effectiveness, Feasibility, and Applications},'' \emph{Applied Sciences}, vol.~14, p. 5889, 07 2024.

\bibitem{guarnera2024mastering}
L.~Guarnera, O.~Giudice, and S.~Battiato, ``{Mastering deepfake detection: A cutting-edge approach to distinguish GAN and diffusion-model images},'' \emph{ACM Transactions on Multimedia Computing, Communications and Applications}, vol.~20, no.~11, pp. 1--24, 2024.

\bibitem{Borzi_2022_CVPR}
S.~Borz{\`\i}, O.~Giudice, F.~Stanco, and D.~Allegra, ``Is synthetic voice detection research going into the right direction?'' in \emph{Proceedings of the IEEE/CVF Conference on Computer Vision and Pattern Recognition (CVPR) Workshops}, June 2022, pp. 71--80.

\bibitem{nagarsheth2017synthetic}
P.~Nagarsheth, X.~Song, J.~Hines, and J.~H. Hansen, ``Detection of synthetic speech for spoofing attack using frame-level deep features,'' in \emph{Proc. INTERSPEECH}, 2017, pp. 706--710.

\bibitem{tak2021rawnet2}
H.~Tak, J.~Patino, M.~Todisco, A.~Nautsch, N.~Evans, and A.~Larcher, ``End-to-end anti-spoofing with rawnet2,'' 2021.

\bibitem{wang2023to}
C.~Wang, J.~Yi, J.~Tao, C.~Zhang, S.~Zhang, R.~Fu, and X.~Chen, ``To-rawnet: Improving rawnet with tcn and orthogonal regularization for fake audio detection,'' \emph{arXiv preprint arXiv:2305.13701}, 2023.

\bibitem{lai2023attentive}
Y.~Lai, Y.~Liu, X.~Liang, J.~Zhu, T.~Zhao, S.~Yu, and C.~Wang, ``Attentive filtering network for audio deepfake detection,'' in \emph{Proceedings of the IEEE/CVF Conference on Computer Vision and Pattern Recognition (CVPR)}, 2023, pp. 3849--3858.

\bibitem{janicki2022specrnet}
A.~Janicki and R.~Heusdens, ``Specrnet: A lightweight end-to-end model for audio deepfake detection,'' in \emph{2022 IEEE International Conference on Acoustics, Speech and Signal Processing (ICASSP)}.\hskip 1em plus 0.5em minus 0.4em\relax IEEE, 2022, pp. 3064--3068.

\bibitem{tak2023asvspoof}
H.~Tak, M.~Todisco, X.~Wang, N.~Evans, J.~Yamagishi, and F.~L. Alegre, ``Asvspoof 2021: Automatic speaker verification spoofing and deepfake detection challenge evaluation plan,'' in \emph{Proc. ASVspoof Workshop 2023}, 2023, challenge paper presenting baseline systems using Wav2Vec 2.0, RawNet2 and AASIST with data augmentation.

\bibitem{guo2024wavlm}
Y.~Guo, H.~Huang, X.~Chen \emph{et~al.}, ``Audio deepfake detection with self-supervised wavlm and multi-fusion attentive classifier,'' in \emph{Proc. ICASSP}, 2024.

\bibitem{zhang2024sls}
Q.~Zhang, S.~Wen, and T.~Hu, ``Audio deepfake detection with self-supervised xls-r and sls classifier,'' in \emph{Proc. ACM MM}, 2024.

\bibitem{coletta2025fpm}
E.~Coletta, D.~Salvi, V.~Negroni \emph{et~al.}, ``Anomaly detection and localization for speech deepfakes via feature pyramid matching,'' \emph{arXiv preprint arXiv:2503.xxxxx}, 2025.

\bibitem{kim2024oneclass}
H.~M. Kim, K.~Jang, and H.~Kim, ``One-class learning with adaptive centroid shift for audio deepfake detection,'' in \emph{Proc. Interspeech}, 2024.

\bibitem{chettri2021end}
S.~Chettri and H.~A. Patil, ``End-to-end spoofing detection with rawnet2 on logical access and physical access,'' in \emph{Proc. Odyssey 2020 The Speaker and Language Recognition Workshop}, 2021, pp. 132--138.

\bibitem{fakeorreal}
R.~Reimao and V.~Tzerpos, ``For: A dataset for synthetic speech detection,'' in \emph{2019 International Conference on Speech Technology and Human-Computer Dialogue (SpeD)}, 2019, pp. 1--10.

\bibitem{todisco2019asvspoof}
M.~Todisco, X.~Wang, V.~Vestman, M.~Sahidullah, H.~Delgado, A.~Nautsch, J.~Yamagishi, N.~Evans, T.~Kinnunen, and K.~A. Lee, ``Asvspoof 2019: Future horizons in spoofed and fake audio detection,'' \emph{arXiv preprint arXiv:1904.05441}, 2019.

\bibitem{xie2024codecfake}
Y.~Xie, Y.~Lu, R.~Fu, Z.~Wen, Z.~Wang, J.~Tao, X.~Qi, X.~Wang, Y.~Liu, H.~Cheng, L.~Ye, and Y.~Sun, ``The codecfake dataset and countermeasures for the universally detection of deepfake audio,'' 2025.

\bibitem{zhang2021deep}
Y.~Zhang, H.~Xie, J.~Xing, X.~Zhen, Z.~Wang, and F.~Liu, ``A survey on deepfake detection: fundamentals, current challenges, and future trends,'' \emph{Computers \& Security}, vol. 112, p. 102494, 2021.

\bibitem{lin2017focal}
T.-Y. Lin, P.~Goyal, R.~Girshick, K.~He, and P.~Doll{\'a}r, ``Focal loss for dense object detection,'' in \emph{Proceedings of the IEEE International Conference on Computer Vision (ICCV)}, 2017, pp. 2980--2988.

\bibitem{sokolova2009systematic}
M.~Sokolova and G.~Lapalme, ``A systematic analysis of performance measures for classification tasks,'' \emph{Information Processing \& Management}, vol.~45, no.~4, pp. 427--437, 2009.

\bibitem{martin1997det}
A.~Martin, G.~Doddington, T.~Kamm, M.~Ordowski, and M.~Przybocki, ``The det curve in assessment of detection task performance,'' in \emph{Proc. Eurospeech}, 1997.

\bibitem{negroni2024moe}
V.~Negroni, D.~Salvi, A.~Ilic~Mezza \emph{et~al.}, ``Leveraging mixture of experts for improved speech deepfake detection,'' in \emph{Proc. ICASSP}, 2025.

\end{thebibliography}
}
\end{document}